\documentclass{article}
\usepackage[a4paper, total={6in, 8in}]{geometry}

\title{Cybersecurity Awareness}
\author{Jason R.C. Nurse\\
University of Kent, UK\\
j.r.c.nurse@kent.ac.uk}

\date{}

\begin{document}

\maketitle

\vspace{2em}
\noindent{\footnotesize{Nurse J.R.C. (2021) Cybersecurity Awareness. In: Jajodia S., Samarati P., Yung M. (eds) Encyclopedia of Cryptography, Security and Privacy. Springer, Berlin, Heidelberg. https://doi.org/10.1007/978-3-642-27739-9\_1596-1}}

\section{Synonyms}
Security awareness, Information security awareness, Cyber awareness.

\section{Definition}
Cybersecurity awareness can be viewed as the level of appreciation, understanding or knowledge of cybersecurity or information security aspects. Such aspects include cognizance of cyber risks and threats, but also appropriate protection measures.   

\section{Background}
Information and communication technologies have become a core part of society. These drive innovation and industry, and are increasingly used to support governments and interaction between workers and members of the public. As these technologies have become more prominent, so too have the threats against them. Cyber-attacks, i.e., threats that leverage cyberspace, in particular, are constantly growing as attackers, hackers, fraudsters and other malevolent actors seek gains in finances, power, influence and recognition.

Cyber-attacks can take many forms, and exploit a wide series of vulnerabilities. For instance, hackers may solicit the services of a botnet to conduct a denial-of-service attack on websites, as in the case of the Mirai botnet attack in 2016. While attacks that target technology can reap substantial rewards, attacks against individuals (at work or at home) are significantly more common and arguably much more effective. Phishing, for instance, is widely regarded as the most dangerous threat to users and organizations online, with some estimates suggesting it accounts for up to 90\% of all security breaches. Equally, the reality of modern-day technology systems means that users can make small mistakes which have far-reaching impacts. For example, a government employee may mistakenly leave a USB drive contain millions of personal data records on the train during their commute, or staff at a healthcare provider may, in error, email sensitive patient records to the wrong address. Human error --- either due to unintentional actions or badly designed human-computer interfaces --- has long been an important concern both in the safety and security fields.

As cyber-attacks targeting users and concerns pertaining to human errors within security have risen, so too have the range of solutions to address them. A leading area of practice and research that has focused on these human aspects in security is that of cybersecurity awareness. 

\section{Theory and Application}
To understand cybersecurity awareness, it is important to first reflect on the topics of information security awareness and cyberspace. Information security awareness as a concept defines an individual’s or group’s understanding of information security, including risks and threats to information, and methods that could be applied to protect such information~\cite{kruger2006prototype}. As technology has advanced, cyberspace – and the online environment – has become more prominent and features in almost every area of modern-day life. To accommodate for this progression, there has been a gradual movement away from information security and towards cybersecurity, both in general and with respect to awareness. Cybersecurity extends the traditional focus on information security, and considers the vast variety of new devices (e.g., laptops, mobile phones, Internet of Things devices and smart tech) that can be exposed to threats, and the methods needed to protect them. Cybersecurity awareness builds on this foundation, to consider the extent to which users are cognizant regarding this wide range of aspects.

Two important questions when discussing cybersecurity awareness are: what is a ‘good’ level of awareness and who is responsible for facilitating it. One way in which this can be explored is to examine the context. In a work context, there are a range of threats including social engineering, phishing, malware, fraud, human error, and insider threat. An appropriate awareness level in this environment would therefore mean having a foundational understanding of these cyber risks, the techniques criminals use to craft their attacks (e.g., fake branding, exploiting cognitive biases~\cite{nurse2018cybercrime}), and the ways in which protection methods can be applied; or, who to speak to for further support. Here, businesses take the responsibility for increasing staff awareness given that successful attacks can directly harm the organization. 

At home and in the public domain, the threats are largely identical but the responsibility and support structures are less present. Governmental and non-governmental organization (NGO) cybersecurity awareness campaigns have attempted to tackle this issue but the lack of direct contact with individuals, and low appeal of some campaigns (which are typically standalone websites, media videos, or posters in public places) often results in poor uptake. This has generally meant that the standard level of cybersecurity awareness in the public is low, and the perception of ‘good’ security hygiene at home is lower than at work. This reality has been a challenge for governments as citizens are constantly falling victim to scams and attacks, and for industry given the increase in remote and home working.

While the need for higher levels of cybersecurity awareness in users of technology to address deliberate and unintentional threats is undisputed, what is less clear is the how awareness is best achieved. Traditional approaches rely on directed reading, posters, or online or in-class training. In organizations these methods are the most common and training, in particular, is often mandated as a part of yearly exercises or refresher courses. Directed reading involves setting specific reading tasks about good security behavior and hygiene; such as how to choose a strong password and being careful of suspicious emails. Posters typically seek to be more eye-catching, delivering short, informative messages. Examples of poster slogans are: “Don’t get hooked. Think before you click.”, “Security. We're all in this together.”, “Storage device. It’s small in size, but big in risk.” and “Passwords are like underwear… Change them often and never share with anyone.”. Training is more directed and can occur in a class environment or through an online system. These often have tests at the end to check understanding and knowledge about the topics discussed. 

One problem with traditional approaches is that they can fail the adequately engage users and therefore, may rarely achieve any long-term behavior change. Directed reading can be tedious, posters may be ignored over time, and class sessions can be viewed as a chore.  Newer approaches have emerged to address these issues in improving a user’s cybersecurity awareness. Security gamification and simulations are examples. Security gamification aims to use elements and principles of game design to make security concepts more engaging and thereby, to better support user awareness of security aspects. Simulations use virtual and physical environments to train and test users. This can range from simulated phishing attacks where users are sent fake phishing emails by a trusted entity (e.g., an employer) and their response is monitored, to 2D or 3D environments where users play real or fictional characters involved in security-related tasks. A structured analysis and discussion on these methods and the extent that users prefer them can be found in Abawajy~\cite{abawajy2014user}.

\section{Cybersecurity Awareness in Practice}
Most organizations now have cybersecurity awareness campaigns of some sort which implement a selection of the approaches mentioned earlier. Online training, simulations and posters are the most commonly deployed, and draw on internal expertise as well as external software training packages. Third-party suppliers have actually become much more popular as businesses look to access experience and knowledge that they lack internally. For instance, many of these suppliers employ psychologists, sociologists, behavioral scientists and marketing professionals; all working towards developing effective awareness programs. Unfortunately, given the financial value of such offers, it is challenging to find citable examples online of effective campaigns in practice.

One set of organizations that have struggled more than most in terms of incorporating awareness efforts have been Small-to-Medium-Sized Enterprises/Businesses (SMEs/SMBs). These organizations have traditionally lagged behind in their resource (time, finances, etc.) investments in security, so it is not surprising when focused on awareness. Even though dedicated support avenues exist in some countries~\cite{bada2019developing}, uptake is a real issue and even in cases where campaigns are launched, they tend to be too basic to result in a tangible, long-term impact. 

On the other hand, campaigns that target the public – and the home user – are widely accessible and may have more success. In the UK, two main campaigns exist; Cyber Aware (UK Government led) and Get Safe Online (which represents a private/public sector partnership). In the USA, the largest campaign is Stop Think Connect. This campaign, which is led by the Anti-Phishing Working Group (APWG) and National Cyber Security Alliance (NCSA), has also been deployed in multiple countries and organizations across the world including in Spain, Japan, India, Bolivia and Jamaica. The guidance provided through these campaigns covers a variety of topics including tips for protecting against online threats, blogs, resources for parents, kids and businesses, and ways to get involved. 

There are also other focused events dedicated to raising awareness of the public and organizations. In Europe, October is noted to be European Cyber Security Month. Similarly, October is National Cybersecurity Awareness Month in the USA and in the Organization of American States (OAS). Lastly, there is also now a Safer Internet Day (SID) on 11 February every year, particularly targeted towards keeping the internet safe for children and young people. This provides a range of content to educate, help, and build the confidence of individuals in protecting themselves online. Currently, there are events and activities in over 170 countries that support the SID. This, and the other efforts highlighted above, clearly highlight the efforts being placed into developing an internet that is safe for all, from businesses to children.

\section{Open Problems and Future directions}
While there have been numerous developments over the last two decades, there are still a plethora of outstanding challenges facing the field of cybersecurity awareness. The most significant pertains to the question of how this awareness is best implemented such that long-term behavior change is achieved. Each of the approaches and campaigns listed have their advantages but it is unclear in which scenarios, contexts or environments they may work the best. Moreover, existing research~\cite{bada2015cyber} has defined factors (e.g., personal factors, culture, persuasive techniques) which significantly impact behavior change, but many of these still seem to be overlooked in current awareness raising efforts, both in the work and public spheres. This is in favor of quick or flashy approaches which lack an appropriate psychological and sociological basis and thus, do not appreciate how users think, interact, make decisions, or engage with security. 

There are also other issues that can arise in the pursuit of increased awareness. For instance, a simulated phishing exercise, where employees are phished by their employers, could end badly. Employees may become distrustful in their employer, or if they are reprimanded for failing phishing tests, could get frustrated and eventually harm (i.e., become an insider threat) or leave the company. This highlights the balance that employers must maintain in trying to protect the business but also maintain a good moral; thus, avoid a blame culture and supporting a Just culture. Future work needs to better appreciate the links between the aforementioned fields especially as cybercriminals use human weaknesses to base their attacks~\cite{nurse2018cybercrime}). These fields can provide useful insights into how to build truly effective awareness campaigns which also can build the wider security culture of a workplace or public environment. 

A second notable challenge relates to defining rigorous metrics for cybersecurity awareness. To adequately assess whether approaches are working, SMART (Specific, Measurable, Attainable, Relevant and Time-based) objectives need to be set. Measures need to go beyond the number of phishing emails clicked/reported and number of poster/page views, to consider underlying factors which better capture user understanding about cyber risks and protective measures. This is particularly important given how difficult it is to demonstrate good return on security investments. 

Finally, research and practice need to appreciate that today’s world is fast-paced and technology, as well as cyber-attacks, are constantly evolving. Cybersecurity awareness approaches are often static and do not accommodate for this changing threat landscape or the innovative nature of cybercriminals. Research is required into adaptive frameworks and platforms for cybersecurity awareness that are flexible and can quickly be updated, to ensure that users stand a chance to protect themselves against attackers. 

\section{Cross-References}
\begin{itemize}
    \item Information Security Culture
    \item Risk Communication
    \item Security Nudges
    \item Persuasive Security Messages
    \item Security Fatigue
    \item Psychology of Cybercrime
\end{itemize}

\bibliographystyle{plain}
\bibliography{main}

\end{document}